\documentclass[a4paper,11pt]{article}
\usepackage{aaskaiid}
\usepackage{orcidlink}

\title{Measuring High-Energy Cosmic Particles with the SKA}
\ShortTitle{Detecting High-Energy Cosmic Particles}

\author[a,d]{Tim~Huege\orcidlink{0000-0002-2783-4772}}
\author[e,j]{Katharine~Mulrey\orcidlink{0000-0001-8026-8020}}
\author[b]{Sjoerd~Bouma\orcidlink{0000-0002-6959-2302}}
\author[c]{Justin~Bray\orcidlink{0000-0002-0963-0223}}
\author[d,e]{Stijn~Buitink\orcidlink{0000-0002-6177-497X}}
\author[d,e]{Arthur~Corstanje\orcidlink{0000-0001-5992-6228}}
\author[d]{Vital~De Henau\orcidlink{0009-0003-0337-3558}}
\author[f]{Edwin~Dickinson\orcidlink{0000-0003-0834-4708}}
\author[g,h]{Brian~Hare\orcidlink{0000-0001-5138-1235}}
\author[i,q]{Haoning~He\orcidlink{0000-0002-8941-9603}}
\author[e,d,j]{J\"org~H\"orandel\orcidlink{0000-0001-6604-547X}}
\author[f]{Clancy~James\orcidlink{0000-0002-6437-6176}}

\author[b]{Philipp~Laub\orcidlink{0009-0003-2617-9109}}
\author[i]{Xingyu Li\orcidlink{}}
\author[g,h]{Marten Lourens\orcidlink{0009-0006-3640-1043}}

\author[a]{Hermann-Josef~Mathes\orcidlink{}}
\author[b,k]{Anna~Nelles\orcidlink{0000-0002-1720-6350}}
\author[a]{Subhadip~Saha\orcidlink{0000-0003-2435-8317}}
\author[r]{Felix~Schl\"uter\orcidlink{0000-0002-5545-4363}}

\author[g]{Olaf~Scholten\orcidlink{0000-0003-3649-1254}}
\author[c]{Ralph~Spencer\orcidlink{0009-0009-6015-1787}}
\author[h]{Christopher~Sterpka\orcidlink{0000-0001-8217-0836}}
\author[b]{Karen~Terveer\orcidlink{0009-0002-9594-0419}}
\author[l]{Satyendra~Thoudam\orcidlink{0000-0002-7066-3614}}
\author[m]{Gia~Trinh\orcidlink{0000-0002-5352-5092}}
\author[h]{Paulina~Turekova\orcidlink{0009-0006-1262-7507}}
\author[a]{Darko~Veberic\orcidlink{0000-0003-2683-1526}}
\author[a]{Keito~Watanabe\orcidlink{0000-0003-0599-4035}}

\author[n,o]{Chao~Zhang\orcidlink{0000-0001-9366-0056}}
\author[p]{Pengfei~Zhang\orcidlink{0000-0002-6855-5315}}
\author[i]{Yi~Zhang\orcidlink{0000-0001-6223-4724}}
\ShortName{T. Huege, K. Mulrey et al.} 

\newcommand{\affilASTRON}{Netherlands Institute for Radio Astronomy (ASTRON), Dwingeloo, The Netherlands}
\newcommand{\affilCanTho}{Physics Education Department, School of Education, Can Tho University, Campus~II, 3/2 Street, Ninh Kieu District, Can Tho City, Viet Nam}
\newcommand{\affilCurtin}{International Centre for Radio Astronomy Research, Curtin University, Bentley, 6102, WA, Australia}
\newcommand{\affilDESY}{Deutsches Elektronen-Synchrotron DESY, Platanenallee~6, 15738 Zeuthen, Germany}
\newcommand{\affilErlangen}{Erlangen Centre for Astroparticle Physics, Friedrich-Alexander-Universit\"at Erlangen-N\"urnberg, 91058 Erlangen, Germany}

\newcommand{\affilGroningen}{Kapteyn Astronomical Institute, University of Groningen, P.O.~Box 72, 9700 AB Groningen, Netherlands}
\newcommand{\affilHefei}{School of Astronomy and Space Science, University of Science and Technology of China, Hefei 230026, China}

\newcommand{\affilKeyNanjing}{Key Laboratory of Modern Astronomy and Astrophysics, Nanjing University, Ministry of Education, Nanjing 210023, China}
\newcommand{\affilKIT}{Institut f\"ur Astroteilchenphysik, Karlsruhe Institute of Technology (KIT), P.O.~Box 3640, 76021 Karlsruhe, Germany}
\newcommand{\affilKhalifa}{Department of Physics, Khalifa University, P.O.~Box 127788, Abu Dhabi, United Arab Emirates}
\newcommand{\affilManchester}{Jodrell Bank Centre for Astrophysics, Department of Physics and Astronomy, University of Manchester, Manchester M13 9PL, UK}

\newcommand{\affilNanjing}{School of Astronomy and Space Science, Nanjing University, Nanjing 210023, China}
\newcommand{\affilNijmegen}{Department of Astrophysics/IMAPP, Radboud University Nijmegen, P.O.~Box 9010, 6500 GL Nijmegen, The Netherlands}
\newcommand{\affilNikhef}{Nikhef, Science Park Amsterdam, 1098 XG Amsterdam, The Netherlands}
\newcommand{\affilPurpleMt}{Key Laboratory of Dark Matter and Space Astronomy, Purple Mountain Observatory, Chinese Academy of Sciences, No.~10 Yuanhua Road, Nanjing, China}
\newcommand{\affilULB}{Universit\'e Libre de Bruxelles, Science Faculty CP230, B-1050 Brussels, Belgium}
\newcommand{\affilVUB}{Inter-University Institute For High Energies (IIHE), Vrije Universiteit Brussel (VUB), Pleinlaan 2, 1050 Brussels, Belgium}
\newcommand{\affilXidian}{School of Electronic Engineering, Xidian University, No.2 South Taibai Road, Xi'an, China}

\affiliation[a]{\affilKIT}
\affiliation[b]{\affilErlangen}
\affiliation[c]{\affilManchester}
\affiliation[d]{\affilVUB}
\affiliation[e]{\affilNijmegen}
\affiliation[f]{\affilCurtin}
\affiliation[g]{\affilGroningen}
\affiliation[h]{\affilASTRON}
\affiliation[i]{\affilPurpleMt}
\affiliation[j]{\affilNikhef}
\affiliation[k]{\affilDESY}
\affiliation[l]{\affilKhalifa}
\affiliation[m]{\affilCanTho}
\affiliation[n]{\affilNanjing}
\affiliation[o]{\affilKeyNanjing}
\affiliation[p]{\affilXidian}
\affiliation[q]{\affilHefei}
\affiliation[r]{\affilULB}

\emailAdd{tim.huege@kit.edu, k.mulrey@astro.ru.nl}

\abstract{The origin of high-energy cosmic rays remain one of astrophysics' greatest unsolved mysteries. SKA-Low will be able to measure air showers initiated by cosmic rays with unprecedented precision in the energy range between $10^{15}-10^{18}$~eV, covering the critical transition region between Galactic and extragalactic sources. SKA-Low's densely instrumented core and broad bandwidth will allow for measurements of individual air showers with a level of detail unmatched by any existing or planned detector. The depth of shower maximum, the primary mass-sensitive observable, will be reconstructed with a resolution of better than 8~g/cm$^2$, a significant improvement over existing methods. Additionally, new reconstruction methods are expected to enable full air shower development reconstruction across a wide energy range, reaching down to PeV levels. At these energies, efficient photon/hadron separation may offer an opportunity to measure PeV gamma-ray air showers. Furthermore, SKA-Low opens a window into studying high-energy hadronic interactions, including via the unique channel of anomalous air showers. This combination of measurements provides a unique opportunity to investigate the origins and physics of high-energy cosmic rays.

A dedicated particle detector array will provide triggered readout of raw antenna-level voltage buffers, enabling fully commensal cosmic-ray observations alongside regular operations. Here we outline our science case and discuss the observational strategy, signal properties and detector design underpinning these measurements. We also summarize the accompanying book chapters, which address composition measurements in the Galactic-to-extragalactic transition region, next-generation interferometric reconstruction techniques, hadronic interaction physics through anomalous air showers, the prospects for detecting PeV gamma-rays from Galactic sources, and the related project of imaging lightning using SKA-Low.}

\begin{document}
\maketitle

\section{Introduction}

Cosmic rays, consisting of ionized atomic nuclei, are the most energetic particles in the universe, with energies spanning over ten orders of magnitude and reaching beyond $10^{20}$~eV. Such extreme energies can only be achieved in highly energetic astrophysical environments that combine strong magnetic fields, large spatial scales, long lifetimes, and high energy densities~\citep{Hillas:1984ijl}. Studying cosmic rays therefore provides a unique probe of the most extreme physical processes in the universe. Yet, despite a century of observations, the origin of high-energy cosmic rays remains unresolved.

Making progress in this field requires measuring significant numbers of cosmic rays and reconstructing their mass (or charge) and energy, as the evolution of cosmic ray composition with energy provides some of the strongest clues about their origin.  At high energies (above $\sim 10^{15}$~eV) cosmic rays can only be detected indirectly via the ``air showers" of secondary particles that are generated when they interact in the atmosphere. Air showers emit very short, non-repeating radio pulses with a duration of order nanoseconds, measurable in the $10\textrm{s} - 100\textrm{s}$~ MHz band. These radio pulses, illuminating a footprint on the ground with a diameter of order 100s of metres,  contain valuable information about the air shower development and therefore the cosmic ray that initiated the shower. The core of SKA-low will be the ideal instrument with which to detect radio emission from air showers, with the large bandwidth, dense antenna spacing, and large number of antennas providing unprecedented precision. In this chapter, we provide an overview of high-energy cosmic particle detection with SKA-Low, first describing the general approach and then shortly describing the other book chapters pertaining to high-energy cosmic particle detection.

\subsection{Open questions in cosmic ray science}

\begin{figure}
    \centering
	\includegraphics[width=0.8\columnwidth]{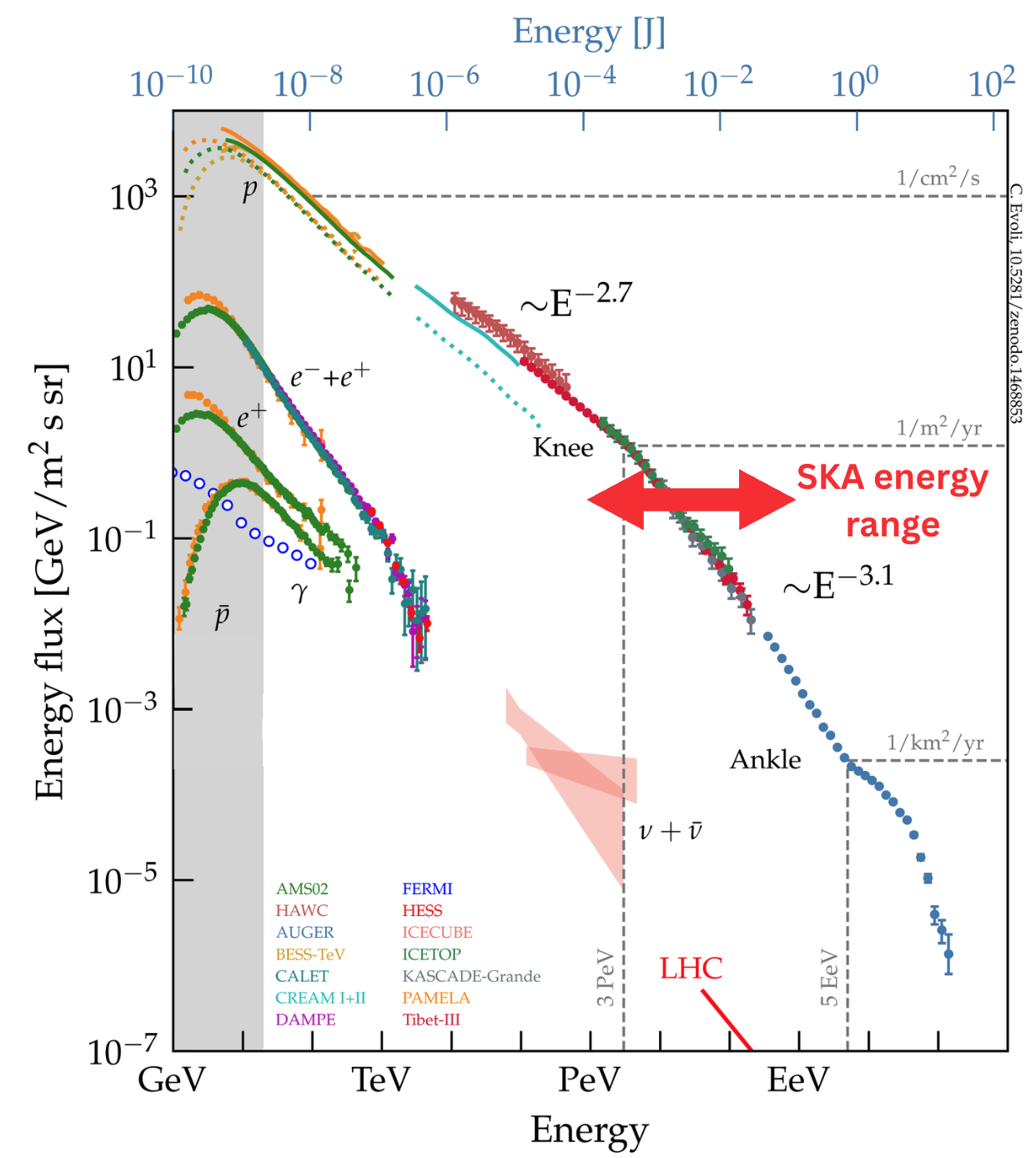}
    \caption{The cosmic-ray energy spectrum, adapted from~\cite{evoli2018_cosmicrayspectrum}. Also shown are the gamma-ray and neutrino fluxes, together with the LHC center-of-mass energy. The arrow indicates the energy range accessible to the SKA.}
    \label{fig:energy_spectrum}
\end{figure}

The origin of high energy cosmic rays is one of astrophysics' greatest unsolved mysteries. Because cosmic rays are charged nuclei, their trajectories are deflected by Galactic and extragalactic magnetic fields, removing directional information about their sources. Instead, source characteristics must be inferred from the energy-dependent composition of cosmic rays, i.e., the relative abundance of nuclei with different masses and charges. The maximum achievable energy depends on the size, magnetic field strength, and lifetime of the acceleration region, as well as the charge of the accelerated particle. Since both acceleration and propagation are governed by particle rigidity, set by charge and energy, the composition encodes the conditions at the source and during propagation. Additionally, particular sources can leave a unique composition fingerprint on the cosmic rays they accelerate. Figure~\ref{fig:energy_spectrum} shows the cosmic-ray energy spectrum as measured with several experimental efforts worldwide.

The cosmic ray energy spectrum spans many orders of magnitude in energy and follows an approximate  $E^{-2.7}$ power-law, but with distinct features that provide important clues about the origin of high-energy particles. Around $4\times10^{15}$~eV, the spectrum steepens in what is called the knee, generally interpreted as the limit of Galactic accelerators such as supernova remnants, where magnetic confinement and shock acceleration reach their maximum effectiveness. A second steepening, the second knee, appears near $10^{17}- 10^{18}$~eV, often linked to the diminishing contribution of heavier Galactic nuclei, suggesting a gradual transition from Galactic to extragalactic sources. At even higher energies, around $3\times10^{18}$~eV, the spectrum flattens again in the so-called ankle. This feature is widely interpreted as the point where extragalactic sources begin to dominate the flux, overtaking the Galactic contribution. Thus, the sequence of knee, second knee, and ankle reflects the interplay between different astrophysical accelerators. With the SKA-Low, we will be able to probe the important region between $10^{15}- 10^{18}$~eV, which covers the entirety of the Galactic to extragalactic source transition region.

\subsection{Cosmic-ray air showers}

Cosmic rays at high energies (above $10^{15}$~eV) must be detected using the air shower of secondary particles that develops when the primary cosmic ray interacts in the Earth's atmosphere. 
The development of an air shower, in particular the total number of electrons and positrons as a function of traversed atmospheric column density (so-called ``depth''), is related to the mass of the primary cosmic ray. The depth of shower maximum $\textrm{X}_{\textrm{max}}$, is the reconstructible quantity that is the most sensitive to the mass of the primary particle: cascades initiated by lighter particles penetrate deeper into the atmosphere and exhibit stronger fluctuations. 

There are a number of approaches to detecting and reconstructing cosmic-ray events. Air shower particles themselves can be measured at ground level using particle detectors. The number of particles, as well as the types of particles (for example, electrons and muons) and their spatial distributions contain information about shower development. The spacing of the particle detectors and the overall ground area covered determines the air shower energy that can be probed. The KASCADE-Grande experiment~\citep{Brancus:2005ht} had about $0.4$~km$^2$ effective area and measured cosmic rays in the PeV region, whereas the Pierre Auger Observatory covers $3,000$~km$^2$ and is optimized to detect the highest-energy cosmic rays~\citep{PierreAuger:2015eyc}. Interpretation of the particle detector data requires Monte Carlo simulations of the air showers. These rely on hadronic interaction models -- which are known to not fully describe the measurements \citep{Albrecht:2021cxw}. Air showers can also be detected using fluorescence light that is generated in the atmosphere as the shower passes (Auger:~\cite{Auger2015}, Telescope Array:~\cite{TA_SD2012, TA_FD2012}). This gives a full picture of the shower development, but is primarily applicable to measuring the highest energy air showers and only works for about 15\% of the time as it relies on dark, moonless and cloudless nights. Cherenkov radiation emitted by relativistic charged particles in the shower can also be detected, either at ground level or using imaging atmospheric Cherenkov telescopes, providing complementary information on shower development and primary energy. See, for example, the Tunka-133 array which measures in the energy range $10^{16}-10^{18}$~eV over a geometric area of $\sim3$~km$^2$~(~\citep{Prosin:2014dxa}).


Since the early 2000s, the radio detection technique, in which cosmic rays are measured using the radio emission that is generated as air showers develop, has emerged as an effective method of cosmic ray detection~\citep{Huege:2016veh}.  This emission is produced primarily by the electromagnetic component of the air shower and can be calculated from first principles.  The dominant contribution to the radio emission comes from the geomagnetically induced, time-varying transverse current that develops as the shower evolves in the Earth’s magnetic field. The strength of this emission scales with the absolute value of the geomagnetic field and the sine of the angle between the shower axis and the local geomagnetic field. A secondary contribution comes from emission resulting from the development of a negative charge excess in the shower front due to ionisation of the atmosphere by the passage of the air shower. 
The measured radio signal constitutes an integral over the whole air shower, so measurements can be used to perform calorimetric energy reconstructions (similar to the fluorescence method) of the energy in the electromagnetic cascade of air showers.  The total energy radiated by the air shower in the form of radio emission is known as the radiation energy, and it scales quadratically with the electromagnetic energy of the air shower as the emission is coherent.  This was demonstrated experimentally by the Auger Engineering Radio Array (AERA) and studied in further detail at the LOFAR telescope~\citep{LOFAR:2013jil}. 


\section{Signal properties of air-shower radio emission}

When radio signals from air showers reach the ground, their properties in both time and frequency domains can vary considerably depending on air shower properties and antenna position relative to the ``impact point'' of the shower, where its central axis intersects the ground. The pulses are intrinsically bipolar and last only a few nanoseconds, a timescale that reflects the ``thickness'' of the ``air-shower pancake''. 

\begin{figure}
    \centering
	\includegraphics[width=0.95\columnwidth]{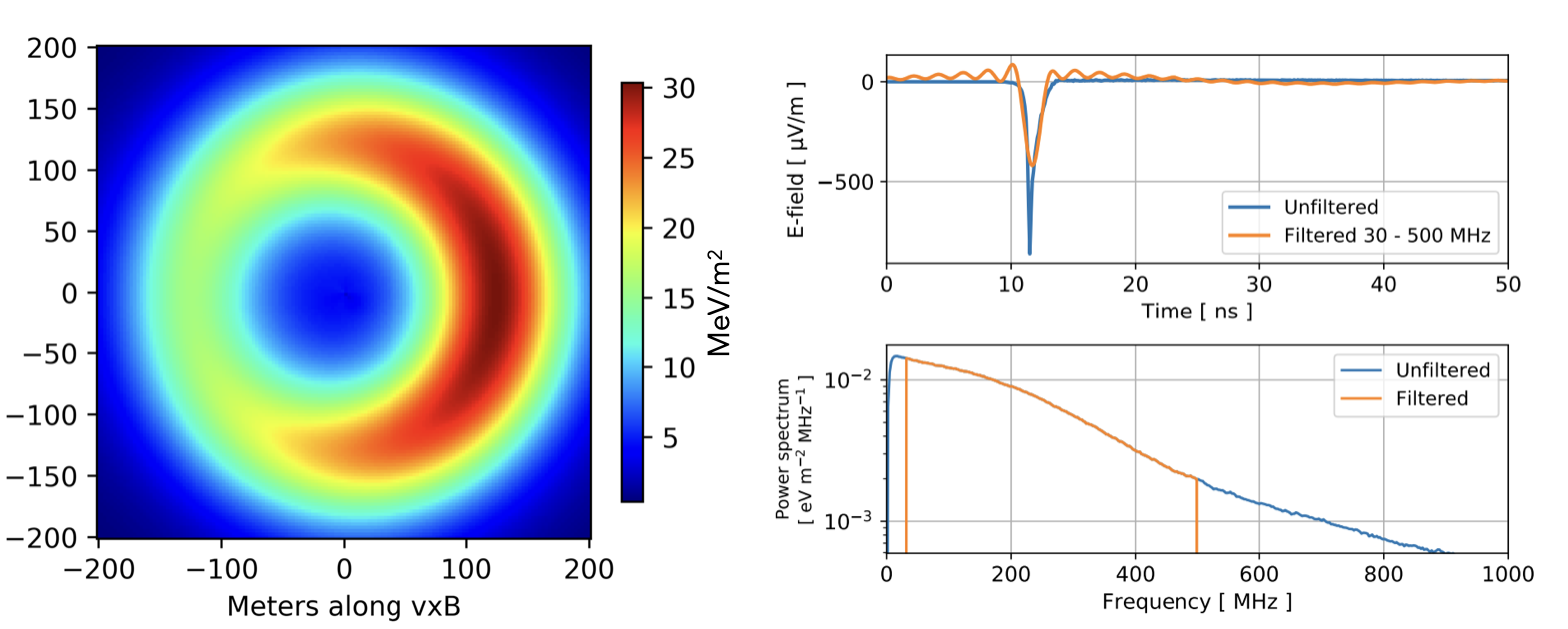}
    \caption{Left: The radio-emission footprint of a typical air shower, filtered to a frequency band of $30-350$\,MHz, indicative of the SKA-Low frequency band of $50-350$\,MHz. Right: Simulated radio pulse for a position inside the Cherenkov cone in the time domain (top) and frequency domain (bottom). Figure adapted from ~\cite{Corstanje:2023vqp}.}
    \label{fig:signal_properties}
\end{figure}

Unfiltered pulses cover a wide frequency band, with the emission usually strongest at tens of MHz frequencies and falling off rapidly, typically exponentially, towards higher frequencies. The steepness of this drop depends on the distance from the ``Cherenkov angle'', the angle at which emission from the complete shower development arrives at the same time due to the interplay of particle velocity and refractive index of the atmosphere through which the radio signals propagate. Starting at the center of the footprint and moving outwards, the high-frequency contribution grows until it peaks near the Cherenkov angle (of order 100\,m for LOFAR or SKA) and then decreases again. The pulses are also highly linearly polarized, with the electric-field orientation determined by the two main radio-emission processes. The dominant geomagnetic mechanism produces polarization in the v×B direction, where v is the shower axis and B is the geomagnetic field, i.e., along the direction defined by the Lorentz force. The charge-excess (Askaryan) component contributes a field directed radially toward the shower axis, typically at the 10--20\% level. The combined polarization pattern changes across the shower footprint, since the interplay between these two mechanisms results in either constructive or destructive interference depending on the location of the observer. The resulting asymmetric radio-emission footprint of a typical shower, illustrating the energy deposit in the form of radio waves per unit area, is shown in the left panel of Figure~\ref{fig:signal_properties}. The right panel shows the time and frequency domain content of a pulse inside the Cherenkov cone.

Beyond the shower properties themselves, the radio emission is also shaped by external factors. For example, the strength and orientation of the local geomagnetic field set the amplitude of the dominant polarization component, the observation altitude affects the stage of shower development sampled by the antennas, and atmospheric conditions, particularly in the case of thunderstorm conditions, influence the radio signal. These influences are now understood well enough to be accounted for in simulations, and modern shower simulation codes reproduce the measured radio footprints, pulse shapes, and polarization patterns with good accuracy across a wide range of shower energies, primary particle types, and detector geometries.

\section{Air-shower detection at the SKA}

\begin{figure}
    \centering
	\includegraphics[width=0.95\columnwidth]{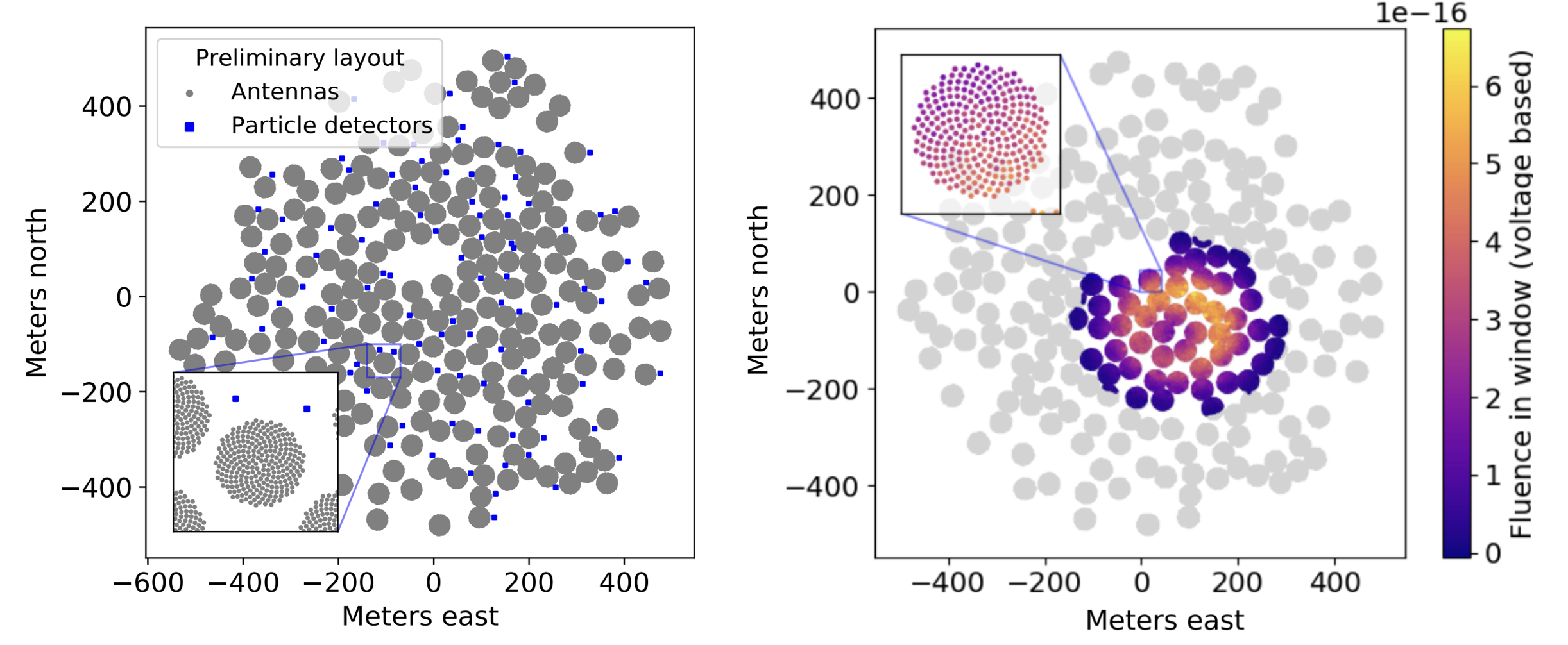}
    \caption{Left: Potential layout of the triggering particle detector array for cosmic ray detection. Right: Radio-emission footprint as it would be measured with SKA-Low. Figure adapted from~\cite{Corstanje:2025wbc}.}
    \label{fig:SKA_layout}
\end{figure}

It was realized already more than ten years ago that SKA-Low will be an extremely powerful air-shower detector \citep{Huege:2014uha,Huege:2015jga}. Compared with other cosmic-ray detectors, it has unique features: while antenna arrays tailor-made for air shower detection, for reasons of cost, typically have antennas deployed on a sparse grid measuring individual air showers with only a handful to a few dozen antennas at a time, the extremely densely instrumented core of SKA-Low will measure individual air showers with thousands of antennas at a time, providing us with an unprecedented wealth of information for each individual measured air shower. Even compared with LOFAR, the densest array with which air-showers have been measured so far, antenna density and uniformity will be on a completely different level. Furthermore, the bandwidth of the SKA antennas, covering the range of $50-350$~MHz is much broader and extending to much higher frequencies than those used in other arrays for air-shower detection, allowing us to extract information from the frequency dependence of the air showers not accessible with other radio detectors. The size of the SKA-Low core region of roughly 1\,km$^2$ determines the energy range that will be accessible for cosmic ray measurements, which will correspond roughly to $10^{15}$ to $10^{18}$\,eV. This is both due to the size of the radio footprint and decreasing statistics at higher energies.
An example of a radio-emission footprint from an air shower of primary energy of $10^{17}$~eV at a zenith angle of 30$^\circ$ as it would be measured with SKA-Low, including realistic noise and antenna response, is shown in the right panel of Figure~\ref{fig:SKA_layout}. The ``fluence in window'' is defined as a time integration of voltage squared in a short time window. The simulations were run using CORSIKA (\cite{Heck:1998vt}) and CoREAS (\cite{Huege:2013vt}) to obtain the radio signals for the site parameters of SKA-Low (altitude 378~m, magnetic field horizontal and vertical component 27.60~$\mu$T and -48.27~$\mu$T respectively).

\subsection{Commensal observations through low-level buffering}

\begin{figure}[h!]
    \centering
	\includegraphics[width=0.95\columnwidth]{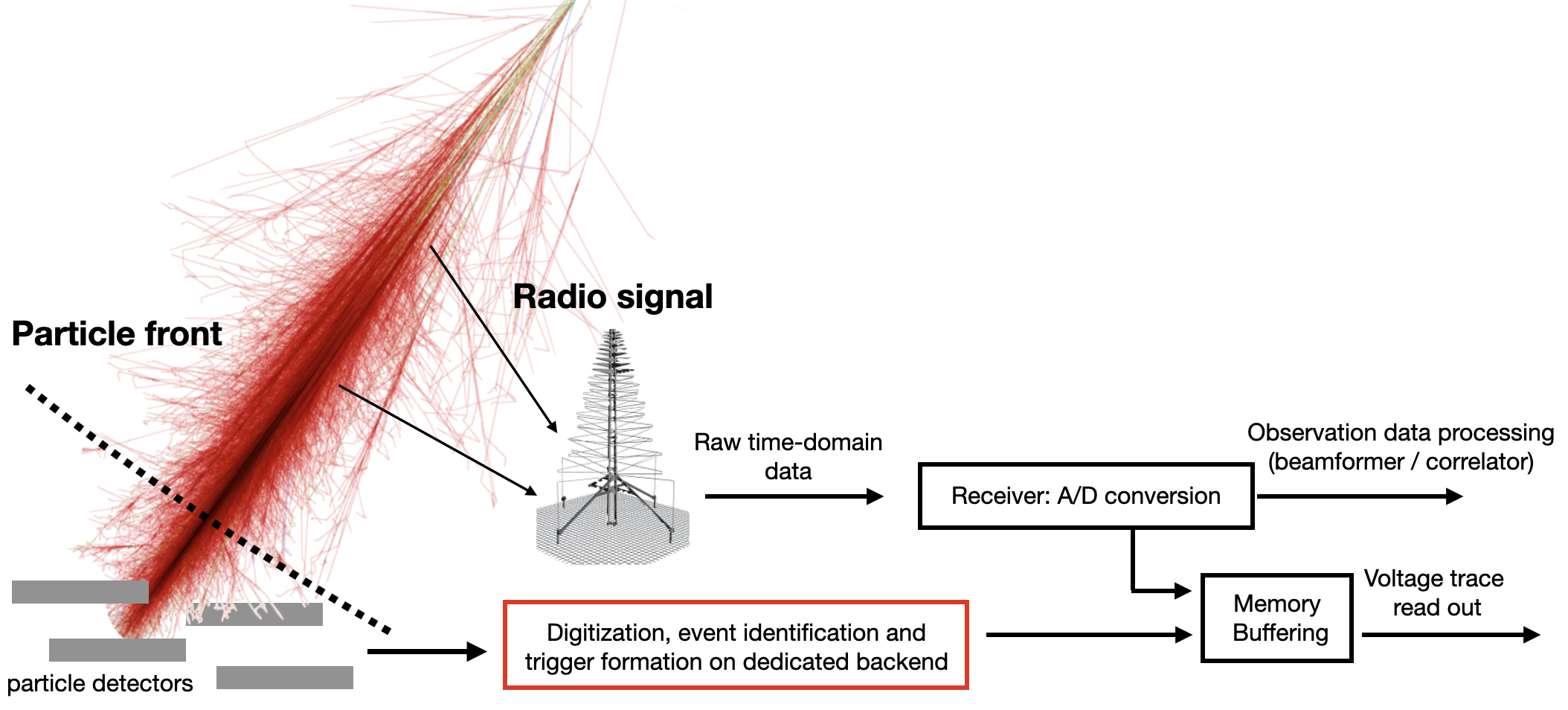}
    \caption{Schematic of cosmic-ray observations at SKA-Low. The red box indicates dedicated electronics to form a cosmic-ray trigger.}
    \label{fig:observation_schematic}
\end{figure}

The detection of air showers initiated by high-energy cosmic particles in the atmosphere above SKA-Low relies on the buffering capability for raw voltage data (before any channelisation or beamforming) of the two polarization channels of each individual antenna. This buffering capability is built in the existing Tile Processing Modules (TPMs). The voltage induced in the individual channels will be digitized and buffered continuously, and upon a ``trigger'', to be delivered by a particle detector array described below, the buffers will be frozen and an approximate microsecond worth of data will be read out from a defined subset of antennas and stored for later, offline analysis (see Figure~\ref{fig:observation_schematic}). The required buffering depth will depend on the latency of the trigger generation. The particle detector trigger will be generated within tens of milliseconds -- additional delays requiring an appropriately large buffer depth might be arising in propagation of the trigger decision through the SKA-Low system. This observing mode will allow fully commensal observations of air showers during ``regular'' astronomical observations, which is a necessary requirement as cosmic-ray studies require the collection of large-statistics data sets, which in turn rely on full-duty-cycle observations.


\subsection{A particle detector array for triggering}

An external cosmic-ray trigger will be generated by installing a particle detector array at the SKA-Low site. As described in the previous section, cosmic-ray air showers produce a particle front that reaches the ground at the same time as the radio signal. The signal from the particle front measured by the particle detector array is distinct and will provide a trigger for cosmic-ray events with shower cores arriving within the SKA-Low core region with 100\% efficiency in the energy range of interest. The high-energy cosmic particles SWG is currently developing the required particle detector design. Prototypes have been tested successfully already within the MWA, located in direct vicinity of the core of SKA-Low \citep{Bray:2020fgw,Dickinson:2026bib}. The most important design goal for these particle detectors is fulfilling the stringent EMI requirements set by the SKAO to ensure that no disturbance of any kind will arise from the particle detector array. This is achieved by avoiding any digital processing, including any clocks, inside the detectors, relying on silicon-photomultipliers instead of classical high-voltage-based photomultipliers, and providing power via DC rather than AC lines. For the backend (power supply, digitization, signal propagation) we will fully rely on SKA-developed and approved hardware. Each detector covers roughly 1\,m$^2$ in area and will be associated with a particular SKA-Low antenna station. A potential layout for the particle detector array is shown in the left panel of Figure~\ref{fig:SKA_layout}.

\section{Book chapters on high-energy cosmic particle detection}

Within the book at hand, the various aspects of high-energy cosmic particle detection and related topics are discussed in detail, showcasing both the state-of-the-art in radio detection of air showers and the new avenues that will open up with the advent of SKA. Here, we give a concise summary of the individual book chapters and how they relate to each other.

\subsection{Chapter: Origins of cosmic rays in the Galactic-extragalactic transition energy range}

A central question in high-energy cosmic-ray research is the maximum energy achievable by Galactic accelerators and the point at which extragalactic sources become the sole contributors. Determining the mass composition of cosmic rays is crucial in this context, as it provides constraints on source models -- both through elemental abundances at the source and through the maximum attained energy, which scales with particle charge. In the chapter by \cite{Corstanje2026.SKA}, it is discussed how the analysis techniques so far developed for LOFAR are expected to perform for SKA-Low. These approaches rely on 1:1 comparisons of measured air-shower radio-emission footprints with simulations using microscopic Monte Carlo codes \citep{Huege:2016veh}. To study their performance for SKA-Low, simulated air-shower radio signals are passed through a detailed SKA-Low detector simulation, complemented by Galactic background emission, and then reconstructed with existing analysis procedures. It is shown that with the increased antenna density and 50--350\,MHz frequency band of SKA-Low, the resolution on the depth of shower maximum -- the primary mass-sensitive observable -- reaches values of 5--8\,g\,cm$^{-2}$, compared with a resolution of 19\,g\,cm$^{-2}$ previously achieved with LOFAR. This is well below the typical separation in $\textrm{X}_{\textrm{max}}$ between different primary mass groups, enabling discrimination not just between light and heavy nuclei but also between finer mass groupings such as helium and CNO. It is also shown that by beam-forming subsets of antennas, the detection threshold can be lowered to energies below $10^{16}$\,eV. Expected event rates are estimated in the energy range of transition from Galactic to extragalactic sources, and the mass-discrimination power is evaluated. Furthermore, there is clearly potential to reconstruct more parameters than just the depth of the air-shower maximum, in particular parameters related to the width and asymmetry of the air-shower evolution profile, which can probe hadronic interaction physics. Finally, the chapter investigates the potential of applying interferometric reconstruction techniques to access the depth of shower maximum, illustrating both the potential but also the challenges involved in applying the currently available techniques to the realistic layout of the SKA-Low core. In summary, this chapter  describes the baseline performance expected to be achievable with SKA air-shower measurements, applying the state-of-the-art analysis techniques proven to work already at LOFAR and other cosmic-ray detectors.

\subsection{Chapter: Interferometric analysis of air-shower radio emission in the near field with an information field theory approach}

The chapter by \cite{Watanabe2026.SKA} investigates how new analysis approaches currently under development in the context of both LOFAR and the SKA high-energy cosmic particle SWG will improve future event reconstruction capabilities. These new approaches, based on ``information field theory'' (a Bayesian-inference–based framework for signal reconstruction that recovers the maximum information content by inferring spatially distributed, field-like quantities), have the potential to overcome the high computational demands, need for simplifying assumptions, and limited use of information contained in the radio signal of the baseline analysis approaches used so far. One approach, developed and applied here for the case of LOFAR data, combines models for the signal wavefront and the spatial distribution of the radio-signal energy fluence (the energy deposit in the form of radio waves per unit area) to reconstruct in particular the energy and depth of shower maximum of individual measured air showers, in a much more computing-efficient way than the 1:1 footprint comparison approach described above. The second approach aims to not only extract the depth of the shower maximum, but the complete longitudinal air-shower evolution profile from the radio measurements with SKA-Low. To this end, it combines information field theory techniques with a fast but accurate forward model for the radio emission from extensive air showers realized with the ``template synthesis approach''~\citep{Desmet:2023shx,Desmet:2025ufy}. In summary, this chapter gives a glimpse of next-generation reconstruction procedures aiming to fully exploit the information content of air-shower radio signal measurements in an efficient form that allows direct inclusion of domain knowledge.

\subsection{Chapter: Anomalous air showers and what they reveal about
hadronic interactions and cosmic-ray masses}

No closed theory exists to describe hadronic interactions in the phase space relevant for extensive air showers. This concerns the energy range (extending beyond the one reachable at the LHC), the very-forward direction, and the specific interaction targets. In fact, all existing hadronic interaction models have shown tension or disagreement with experimental data \citep{Albrecht:2021cxw}. Air-shower measurements with SKA-Low have the potential to test hadronic interaction models in an unprecedented way through the measurement of so-called ``anomalous air showers''. These are air showers in which individual particles carry energy very deeply into the atmosphere so that the resulting air shower rather resembles the superposition of two (or more) ``normal air showers''. In the chapter by \cite{Buitink01.2026.SKA} it is described how such showers can be identified through radio-emission measurements with SKA-Low. They exhibit specific features such as a double-ring structure in the radio-emission footprint as well as interference patterns imprinted on the frequency spectra of the radio-emission pulses. This novel approach, exploitable through the low detection threshold and broad bandwidth achievable only at SKA-Low, has high potential to further our knowledge in hadronic interaction physics, an essential foundation for the interpretation of air-shower measurements not only with the radio detection technique.

\subsection{Chapter: Using SKA-Low to detect PeV gamma-rays from Galactic
Sources}

Identifying PeVatrons is regarded as a central objective of gamma-ray astronomy. These are Galactic astrophysical sources capable of accelerating cosmic rays beyond a PeV ($10^{15}$~eV). Their exact nature remains uncertain, though a few potential candidates have recently been found that just cross the PeV threshold. The SKA-Low, with its vast number of antennas and a core region large enough to provide substantial effective area, may be able to push the energy threshold for detection of gamma-induced air showers low enough to observe PeVatrons. This is a challenging endeavour, as even in case of detectability, the gamma-induced air showers will be overshadowed by a large background from cosmic-ray induced air showers. On the other hand, the very good angular resolution achievable with SKA-Low air shower detection could provide a means to distinguish between hadronic and gamma-induced air showers. That said, triggering with a particle detector will also be difficult at these energies as the purely electromagnetic showers induced by photons exhibit no relevant muonic component and the electromagnetic component will be significantly absorbed in the atmosphere. In the chapter by \cite{Nelles2026.SKA} the opportunities and challenges for detection of air showers induced by high-energy photons are discussed. Gamma-ray detection with the SKA would not only offer a fresh observational approach to investigating Galactic particle accelerators, but could also mark the first successful detection of gamma-ray–induced air showers through radio techniques.

\subsection{Chapter: Unveiling the mysteries of lightning: Exploring its
fundamental physical processes with SKA-Low}

While it may seem initially counterintuitive, the HECP working group also works very closely with the lightning science team. The reason is that both groups use the same level-0 data (single-antenna transient voltage buffers), and similar low-level data processing techniques. This is in addition to the historical reason, that the lightning group grew out of the LOFAR Cosmic Ray group when it was discovered that the electric fields in thunderstorms can impact cosmic-ray propagation~(\cite{Trinh:2016}). 

Lightning is not well understood. We do not understand how lightning initiates, grows through the atmosphere, emits high-energy radiation, or results in innumerable other phenomena that are too many to list~(\cite{Dwyer:2014}). For example, we know the dielectric strength of air is around 3 MV/m at sea-level pressure. However, the electric fields measured in thunderstorms never come close to this level even after accounting for atmospheric pressure~(\cite{Dwyer:2014}). Therefore, lightning must be initiated by some non-conventional breakdown mechanism. Many hypothesis have been proposed e.g.~(\cite{Griffiths:1976}), however none have sufficient experimental evidence to confirm. We do not even understand why lightning emits VHF radio radiation, which is thought to be due to streamer activity that occurs in front of the very hot/conducting channels. But how streamers create VHF is not understood. These questions exist because, among other reasons, lightning is a very challenging phenomena to observe with high precision; it is extremely fast and occurs at random times and locations, behind optically opaque clouds. 

In order to tackle these challenges in lightning science, we will use our experience from working with the LOFAR radio telescope to image lightning with SKA-Low. SKA-Low’s wide bandwidth will allow us to probe different VHF emission mechanisms and create very high precision images. Combined, high spatial precision and wide spectra will allow us to explore the physics of how lightning plasma develops at a level never accomplished before. While much of lightning is very powerful, many key processes (such as initiation) can be very quiet. SKA-Low will also allow for observations of lightning with previously unreached sensitivities, thus enabling us to explore lightning phenomena that have never been observed before. More details of lightning observations with SKA-Low are given in the chapter by~\cite{Hare01.2026.SKA}.

\section{Conclusion}

The core of SKA-Low will be an exceptional detector for extensive air showers induced in the atmosphere by charged cosmic rays, and lightning flashes, and potentially high-energy photons. The Science Working Group on High-Energy Cosmic Particles is actively working on enabling these measurements through the development of an RFI-quiet particle detector array that will enable the triggered readout of antenna-level buffers upon arrival of an air shower. At the same time, efforts are ongoing to predict the expected performance of SKA-Low as an air shower detector. Here classical analysis strategies are further developed and complemented with new approaches that are only feasible with SKA-Low thanks to its uniquely dense antenna configuration and broad bandwidth.

\bibliographystyle{abbrvnat}
\bibliography{chapter} 

\end{document}